\documentclass[useAMS,usenatbib,usegraphicx]{mn2e}
\usepackage{times}

\title[]
  {Kinematic groups beyond the Solar neighbourhood with RAVE}
\author[T. Antoja et al.]
  {
T. Antoja$^1$\thanks{E-mail: antoja@astro.rug.nl},  
A. Helmi$^1$, 
O. Bienayme$^{2}$, 
J. Bland-Hawthorn$^{3}$, 
B. Famaey$^2$, 
K. Freeman$^{4}$,\newauthor 
B. K. Gibson$^5$,
G. Gilmore$^{6}$,
E. K.\ Grebel$^{7}$, 
I. Minchev$^{8}$,
U. Munari$^{9}$,
J. Navarro$^{10}$,\newauthor 
Q. Parker$^{11,12,13}$,
W. Reid$^{14}$,
G. M. Seabroke$^{15}$,
A. Siebert$^{2}$, 
A. Siviero$^{16,8}$, 
M. Steinmetz$^{8}$,\newauthor 
M. Williams$^{8}$,
R. Wyse$^{17}$,
and T. Zwitter$^{18,19}$
\\
$^1$Kapteyn Astronomical Institute, University of Groningen, PO Box 800, 9700 AV Groningen, the Netherlands \\
$^2$Observatoire Astronomique, Universit\'e de Strasbourg, CNRS UMR 7550, Strasbourg, France\\
$^{3}$Sydney Institute for Astronomy, University of Sydney, NSW 2006, Australia\\
$^{4}$RSAA Australian National University, Mount Stromlo Observatory, Cotter Road, Weston Creek, Canberra, ACT 2611, Australia\\
$^5$Jeremiah Horrocks Institute,  University of Central Lancashire, Preston, PR1 2HE, United Kingdom\\
$^{6}$Institute of Astronomy, Cambridge University, Madingley Road, Cambridge CB3 0HA, United Kingdon\\
$^{7}$Astronomisches Rechen-Institut, Zentrum f\"ur Astronomie der Universit\"at Heidelberg, M\"onchhofstr.\ 12--14, 69120 Heidelberg, Germany \\
$^8$Leibniz-Institut f\"ur Astrophysik Potsdam (AIP), An der Sternwarte 16, D - 14482, Potsdam, Germany\\
$^{9}$INAF Osservatorio Astronomico di Padova, Via dell'Osservatorio 8, Asiago, I-36012, Italy \\
$^{10}$Department of Physics and Astronomy, University of Victoria, Victoria, BC V8P 5C2 Canada\\
$^{11}$Australian Astronomical Observatory, Epping, PO Box 296, NSW 1710, Australia\\
$^{12}$Macquarie Research Centre in Astronomy, Astrophysics and Astrophotonics, Macquarie University, NSW 2109, Australia\\
$^{13}$Department of Physics \& Astronomy, Macquarie University, NSW 2109, Australia\\
$^{14}$Dept. Astronomy, Astrophysics and Astrophotonics, Macquarie University, Sydney, Australia\\
$^{15}$Mullard Space Science Laboratory, University College London, Holmbury St Mary, Dorking, RH5 6NT, UK\\
$^{16}$Department of Physics and Astronomy, Padova University, Vicolo dell'Osservatorio 2, I-35122 Padova, Italy\\
$^{17}$Johns Hopkins University, Homewood Campus, 3400 N Charles Street, Baltimore, MD 21218, USA  \\
$^{18}$University of Ljubljana, Faculty of Mathematics and Physics, Ljubljana, Slovenia \\
$^{19}$Center of Excellence SPACE-SI, Ljubljana, Slovenia
}
\date{}

\pagerange{\pageref{firstpage}--\pageref{lastpage}} \pubyear{2012}

\newcommand{\kms}{\rm\,km\,s^{-1}}
\newcommand{\pc}{\rm\,pc}
\newcommand{\kpc}{{\rm\,kpc}}

\def \sun{{_\odot}}
\def \2s {2-$\sigma$ }
\def \3s {3-$\sigma$ }

\def\LaTeX{L\kern-.36em\raise.3ex\hbox{a}\kern-.15em
    T\kern-.1667em\lower.7ex\hbox{E}\kern-.125emX}

\begin{document}

\label{firstpage}

\maketitle

\begin{abstract}
We analyse the kinematics of disc stars observed by the RAVE survey in and beyond the Solar neighbourhood. We detect significant overdensities in the velocity distributions using a technique based on the wavelet transform. We find that the main local kinematic groups are large scale features, surviving at least up to $\sim1\kpc$ from the Sun in the direction of anti-rotation, and also at $\sim700\pc$ below the Galactic plane. We also find that for regions located at different radii than the Sun, the known groups appear shifted in the $v_R$--$v_\phi$ velocity plane. For example, the Hercules group has a larger azimuthal velocity for regions inside the Solar circle and a lower value outside. We have also discovered a new group at $(U, V)=(92, -22)\kms$ in the Solar neighbourhood and confirmed the significance of other previously found groups. Some of these trends detected for the first time are consistent with dynamical models of the effects of the bar and the spiral arms. More modelling is required to definitively characterise the non-axisymmetric components of our Galaxy using these groups.
\end{abstract}


%
\section{Introduction}\label{intro}

The presence of kinematic structures in the velocity distribution of the Solar neighbourhood has been well established through the analysis of Hipparcos and other surveys (\citealt{Dehnen98,Famaey05,Antoja08}, hereafter A08). Some of the structures may be related to orbital effects of the bar \citep{Dehnen00}, spiral arms (\citealt{Antoja11}, hereafter A11) or both (\citealt{Antoja09,Quillen11}, A11). Thus, they can provide constraints on the properties of the non-axisymmetries of the Milky Way (MW) disc. However, the origin of each structure is not yet clear. Some of them could be remnants of disrupted disc stellar clusters or could be associated to accretion events (see review in \citealt{Antoja10}).   

The RAVE (RAdial Velocity Experiment, \citealt{Steinmetz06}) spectroscopic survey provides the means for a giant leap forward in the understanding of the origin of the kinematic groups, allowing us to answer primary questions such as: what are the significant groups in the Solar volume? are they local features? do they change when we move far from the Sun? The RAVE Solar neighbourhood sample contains already more stars than any previously available kinematic data set, such as the Geneva-Copenhagen survey (GCS, \citealt{Holmberg09}). But more importantly, the volume sampled by RAVE is much larger, giving kinematics representative of distant regions of the thin and thick discs. In this Letter, we use the RAVE data to revisit the kinematic groups in the Solar neighbourhood and trace them, for the first time, far from the Sun. 
\vspace*{-0.5cm}

\section{The RAVE data and methods}\label{sample}

\begin{figure}
  \centering
  \includegraphics[width=0.3\textwidth]{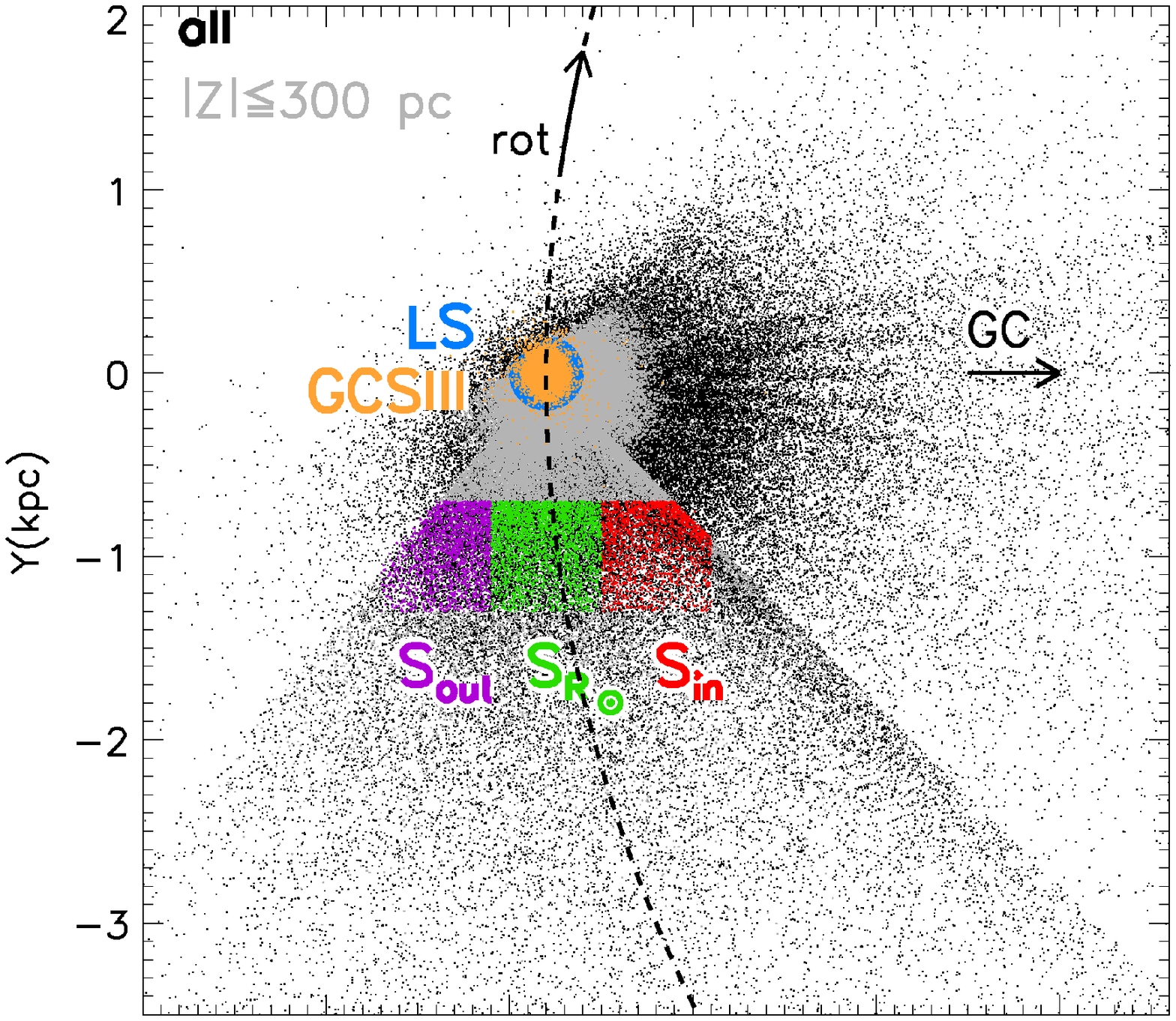}

  \includegraphics[width=0.3\textwidth]{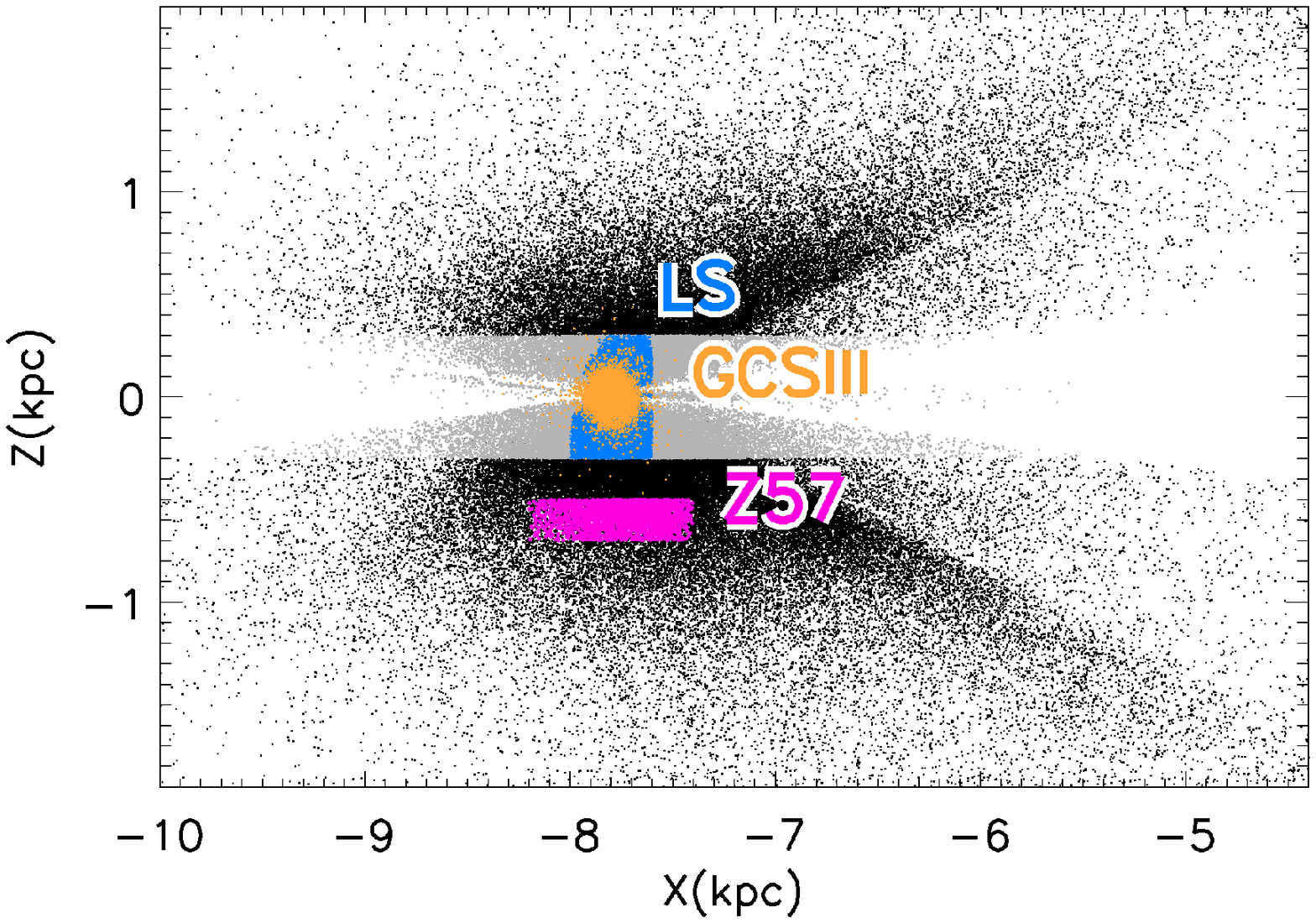}
  \caption{Spatial distribution in $X$-$Y$ and $X$-$Z$ of RAVE data with 6D phase space coordinates (black points). Grey points are stars with $|Z|\leq300\pc$. Orange points are stars in the GCSIII. Other colours are the RAVE subsamples defined in the text. The black dashed line shows the Solar circle.}
  \label{f:XYrave}
\end{figure}

\begin{table}
\caption{Properties of the subsamples defined in the text. For each subsample we give the number of stars, median distance in $\kpc$, median relative error in distance, and median errors in heliocentric velocities (for local subsamples) or in cylindrical velocities (for distant subsamples ) in $\kms$.}           
\label{t:subsamples}      
\centering                          
\begin{tabular}{lrrrrrrrrr}      
\hline
       & \multicolumn{1}{c}{N}& \multicolumn{1}{c}{$\tilde{d}$}&\multicolumn{1}{c}{$\tilde{e}_d$(\%)} & $\tilde{e}_{v_R}$ & $\tilde{e}_{v_\phi}$   & $\tilde{e}_W$\\ \hline 
 GCSIII& 13491& 0.07& 6.&   1. &  1. & 1. \\
     LS& 57178& 0.19&23.&   4. &  4. & 3. \\
    Z57&  2019& 0.65&36.&  11. & 11. & 4. \\
     S$_{\rm{in}}$& 1891 & 1.11&25.&  16.&  11.&  14. \\
{S$_{\rm{R\sun}}$}& 3145 & 0.92&27.&  15.&   4.&  10. \\
     S$_{\rm{out}}$& 2007 & 1.11&29.&   13.&   8.&  13. \\
\hline 
\end{tabular}
\end{table}

RAVE is a magnitude limited multi-fibre spectroscopic survey in the range $9< I< 13$ using the 1.2-m UK Schmidt Telescope of the Anglo-Australian Observatory (AAO). We use here internal data from \citet{Burnett11}, who derived distances for stars in the 2nd and 3rd data releases \citep{Zwitter08,Siebert11b}. Following \citet{Burnett11}, we increase the distances by $10\%$ for hot dwarfs and decrease them by the same factor for giants. The data set includes also proper motions mainly from PPMX \citep{Roser08} and UCAC2 \citep{Zacharias04}, making a sample of 202843 stars with 6D phase space information.
 
The spatial distribution of the sample is shown in Fig.~\ref{f:XYrave} (black points). We use the Galactic Cartesian coordinate system with the Sun at $X=-7.8\kpc$, $Y=0.\kpc$ (2nd model in table 2 of \citealt{McMillan10}) and $Z=0.014\kpc$ \citep{Binney97}. The spatial distribution can be explained by the fact that RAVE is a southern hemisphere survey ($\delta<2^\circ$). The cone through negative $Y$ is due to the additional colour cut $J-K>0.5$ that was applied for latitudes $|b|<25^\circ$ and longitudes $230^\circ<l<315^\circ$ to select giant stars and probe larger distances in the disc. There are 112808 stars with $|Z|\leq300\pc$ (grey points). As a comparison, throughout the paper we will use the GCS sample (hereafter, GCSIII) taken from \citet{Holmberg09}. The main differences between the two data sets are that RAVE is a southern hemisphere survey whereas GCSIII covers the whole sky, that it is much fainter and that it is not restricted to F and G dwarfs. We plot the GCSIII with orange points in Fig.~\ref{f:XYrave}. The volume sampled by the RAVE stars is much larger and the kinematics of the thin disc can be now traced beyond $\sim 1\kpc$ in the direction of anti-rotation. 

We study the kinematics of stars in several spatially distinct samples whose main properties are in Table \ref{t:subsamples}. First, we define the local subsample (LS, blue points in Fig.~\ref{f:XYrave}) as those stars with $|Z|\leq300\pc$ and in-plane distances from the Sun $d\cos(b)\leq 200\pc$. There are 57178 stars in the LS, which is more than four times the number in the GCSIII. Secondly, we select stars in three regions in the plane ($|Z|\leq300\pc$) corresponding to cubic volumes of $0.6\kpc$ size at a distance of $\sim 1\kpc$ from the Sun in the range $-1.3\leq Y<-0.7\kpc$ but at different Galactocentric radius: i) {S$_{\rm{R\sun}}$}: Solar radius $-8.1< X\leq-7.5\kpc$ (green points), ii) S$_{\rm{in}}$: inner radius $-7.5< X\leq-6.9\kpc$ (red points), and iii) S$_{\rm{out}}$: outer radius $-8.7< X\leq-8.1\kpc$ (purple points). Finally, the subsample Z57 (pink points) contains stars in a cylinder with $d\cos(b)\leq 400\pc$ and height below the plane $-700\leq Z\leq-500\pc$. Note that velocity errors in the RAVE subsamples (last columns in Table \ref{t:subsamples}) are dominated by errors in the transverse motion (for which distances and proper motions are needed) as the radial velocity accuracy is considerably better ($e_{v_{los}}\sim2\kms$).

To detect structures in the 2d velocity distributions of these subsamples we use a technique based on the wavelet transform (WT) \citep{Starck02}.
The WT techniques were first applied to detect local streams by \citet{Chereul99}. We use here the {\em \`a trous} algorithm \citep{Starck02}. This consists of first applying a set of smoothing filters with progressively increasing scales to the original data. Then the difference between two successive smoothed planes gives the WT coefficients, which contain the information on the structures present in between the two scales or sizes. For every position in the velocity plane, a WT coefficient equal to 0 corresponds to a constant signal, while a positive (negative) value is associated to overdensities (underdensities). Typical sizes of the kinematic structures are $\sim10 \kms$, so in the absence of errors they will be prominent on maps in between scales of $\sim 6$ and $\sim 15\kms$. However, in practice the velocity errors may determine the scales that will show reliable groups.

Then we search for maxima among the WT coefficients to detect the different groups. Finally, to analyse their significance we use 
the autoconvolution histogram method 
developed in \citet{Slezak93} especially suited for low intensity levels. This evaluates the probability that the WT coefficients are not due to Poisson noise by comparing, in wavelet space, the local distribution of stars to what would result for a uniform distribution. We will consider coefficients that have probability of being real detections $P\ge 99.7\%$ and $95.4\%\le P< 99.7\%$, similar to the \3s and \2s significance levels in the Gaussian case, respectively. 
Note that the computation of the significance refers to that of the wavelet coefficients and takes into account the particular distribution of stars (and not only their number) in the structures, and hence it is not directly comparable with the one obtained by counting stars. Nevertheless, in various tests we have performed, we found that the two significances are similar. Tests with randomised data show that by chance one finds 8\% of the maxima with a 2-sigma significance while 2\% reach a 3-sigma significance level.
To perform the calculations we use the MR software developed by CEA (Saclay, France) and Nice Observatory. An estimate of the number of stars in each group is obtained by counting stars enclosed in a circle with the radius of the minimum scale of the considered range of scales.

\vspace*{-0.7cm}

\section{Revisited Solar volume kinematics}\label{resLS}
\begin{table}
\caption{Kinematic groups detected in the local sample (LS) at scales of $6$-$13\kms$. The columns give: order in WT coefficient, name of the group in previous literature (if any), $U$ velocity of the peak, error in the mean $U$ of the stars in the group, $V$ velocity, error in the mean $V$, approximate number of stars and statistical significance. The last three columns show the equivalent groups for the GCSIII. Velocities are in $\kms$.}
\label{t:groups}
\centering  
    \tabcolsep 3.pt
\begin{tabular}{r l r r r r  r l | r  r r }
\hline
 \multicolumn{8}{l|}{RAVE} & \multicolumn{3}{l}{GCSIII} \\\hline    
 \multicolumn{1}{c}{n.}& Name& $U$& ${e}_U$& $V$& ${e}_V$& \multicolumn{1}{c}{N} &  \multicolumn{1}{c}{sig}& \multicolumn{1}{|c}{n.}&  \multicolumn{1}{c}{N} &  \multicolumn{1}{c}{sig}\\
\hline    
 1&  Coma Berenices             &  -7 & $_{\pm0.1}$ &   -6 & $_{\pm0.1}$ & 2173 & 3  & 4   &  499     & 3   \\
 2&  Hyades                     & -30 & $_{\pm0.2}$ &  -13 & $_{\pm0.1}$ & 1817 & 3  & 1   &  671     & 3   \\
 3&  Sirius                     &   4 & $_{\pm0.1}$ &    4 & $_{\pm0.1}$ & 1558 & 3  & 3   &  465     & 3   \\
 4&  Pleiades                   & -16 & $_{\pm0.1}$ &  -22 & $_{\pm0.1}$ & 1800 & 3  & 2   &  568     & 3   \\ 
 5&  Wolf 630                   &  28 & $_{\pm0.3}$ &  -21 & $_{\pm0.2}$ &  755 & 3  & 10  &  212     & 2   \\
 6&  Hercules I                 & -57 & $_{\pm0.8}$&  -48 & $_{\pm0.7}$ &  300 & 3  &     &          &     \\
 7& Dehnen98                    &  48 & $_{\pm0.6}$ &  -24 & $_{\pm0.4}$ &  352 & 3  & 6   &  145     & 3   \\
 8&  Hercules II                & -28 & $_{\pm0.4}$ &  -50 & $_{\pm0.5}$ &  473 & 2  & 5   &  148     & 3   \\
 9&  $\gamma$Leo                &  56 & $_{\pm0.9}$&    2 & $_{\pm0.3}$ &  193 & 2  &11   &   49     & 2   \\
10&  $\epsilon$Ind              & -81 & $_{\pm1.5}$&  -42 & $_{\pm0.9}$ &  130 & 2  &12   &   51     & 2   \\
11&  $\gamma$Leo                &  68 & $_{\pm1.5}$&    1 & $_{\pm0.4}$ &  111 & 2  &     &        &    \\
12&  NEW                        &  92 & $_{\pm3.5}$&  -23 & $_{\pm1.4}$ &   48 & 3  &     &          &     \\
13&                             &-103 & $_{\pm3.8}$&  -41 & $_{\pm1.7}$ &   51 & 3  &15   &   14       & 2    \\
14&  $\eta$Cep                  & -27 & $_{\pm1.6}$ &  -95 & $_{\pm4.8}$ &   42 & 2  &     &  &  \\
15&                             &  60 & $_{\pm3.3}$ &  -72 & $_{\pm4.7}$ &   29 & 2  &     &  &     \\
16&                             &-122 & $_{\pm8.8}$&  -19 & $_{\pm2.0}$ &   21 & 2  &     &  &    \\
17&                             &  12 & $_{\pm2.1}$ & -108 & $_{\pm4.9}$ &   19 & 2  &     &  &    \\
18&                             & 109 & $_{\pm6.7}$&  -32 & $_{\pm3.6}$ &   17 & 2  &     &  &   \\
19&                             &  27 & $_{\pm3.4}$ & -120 & $_{\pm10.2}$ &   13 & 2  &     &  &   \\
\hline                                   
\end{tabular}
\end{table}

\begin{figure*}
   \centering  \includegraphics[height=0.25\textwidth]{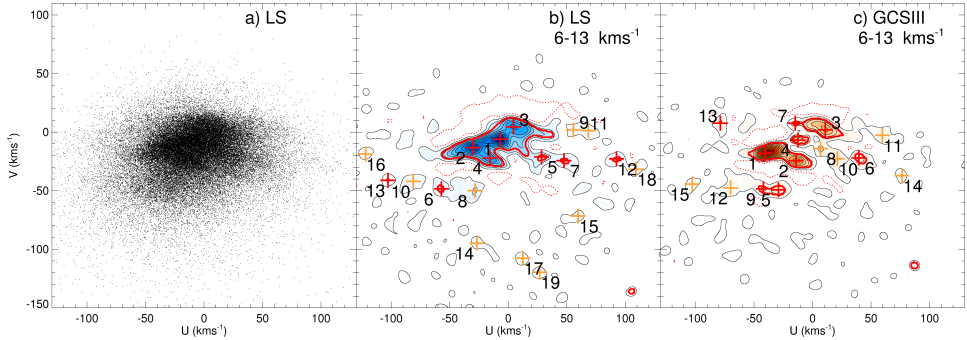}
  \caption{a) Heliocentric velocities of the local sample (LS). b) Velocity structures in the local sample (LS) obtained with the wavelet transform at scales of $6$-$13\kms$. The red solid (dotted) contour shows the \3s confidence limit for positive wavelet coefficients or overdensities (negative coefficients or underdensities). Red (orange) crosses show the maxima of the detected peaks inside the \3s (2-$\sigma$) confidence limit. c) Same as b, but for GCSIII sample.}
  \label{f:UV}%
\end{figure*}

Fig. \ref{f:UV}a shows the velocities of stars in the LS. We use a heliocentric system $U$, $V$, $W$, with the $U$ towards the Galactic Centre. We focus only on the $U$ and $V$ components as the moving groups, except the very young ones \citep{Chereul99}, have almost indistinguishable mean $W$ \citep{Dehnen98,Seabroke07}. Fig. \ref{f:UV}b corresponds to the positive WT coefficients (overdensities) in the LS velocity distribution at scales of $6$-$13\kms$ (where 6 and $13\kms$ are the scales of the smoothing filters used in the computation of the WT). This  distribution is dominated by clear overdensities. Some structures are elongated in $V$ and slightly tilted. We find 9 peaks inside the \3s confidence level region (marked in red) and 10 more that are \2s significant (orange). The peaks are numbered from maximum to minimum WT coefficient and are listed in Table \ref{t:groups}. Fig. \ref{f:UV}c shows the same WT for the GCSIII. We thus find similar significant groups in the two independent samples. But in comparison to previous work (e.g. \citealt{Famaey05}, A08), with RAVE we have better statistics.

The four main groups that we find in the LS (from 1 to 4) are the well-known Coma Berenices, Hyades, Sirius and Pleiades. These groups were also observed in the RAVE sample by \citet{Hahn2011}. Their positions agree well with the GCSIII (separation between maxima $<10\kms$). The only exception is the Hyades stream: the stars of the Hyades open cluster in the GCSIII at ($U$, $V$)=($-43$,$-18$)$\kms$ (absent in the LS) shift the peak to more negative $U$. The different ranking of these 4 groups in the LS and GCSIII may be due to differences in the populations of the samples. 

 The Hercules structure is the elongated feature in the $U$ direction at $V\sim-50\kms$. In the LS, group 6 is its main overdensity. Also groups 8, 10 and 13 (this with higher significance than for the GCSIII) could be associated to it. Notice that the detection of localised maxima in elongated and extended structures such as Hercules may be a bias due to the isotropy of the WT. The red dotted contours between the main four groups and the Hercules structure correspond to a \3s significant underdensity of stars (negative WT coefficients). Another remarkable \3s underdensity is the so-called edge line (\citealt{Skuljan99}) in the upper part of the distribution which is present both in the LS and the GCSIII. 
 
The fifth group in order of WT coefficient (before the Hercules stream) is close to the classic group Wolf 630 studied in \citet{Eggen71}. Group 7 is linked to this group in a branch-like shape (hereafter Wolf elongation). We also find other groups that can be associated to the ones studied by \citet{Eggen71,Eggen96b}. For instance, group 10 is similar to $\epsilon$Indi, group 9 and 11 are close to $\gamma$Leo, and group 14 could be identified with $\eta$Cep.

Group 12 is a new \3s significant group at ($U$, $V$)=($92$, $-22$)$\kms$. It is accompanied by group 18 but only at \2s level. It has around 50 stars but its location in an empty region of the distribution makes it conspicuous. Notice that this is a group on a highly eccentric orbit, moving towards the Galactic Centre at $\sim100\kms$. There are other groups (15-19) yet to be confirmed as they are only \2s significant. Some of them could be tentatively related to other known groups. For instance, groups 14, 17 or 19 could be related to Arcturus at $V\sim-110\kms$ \citep{Eggen96b}. We leave a more thorough comparison to groups in other studies (e.g. \citealt{Klement11,Williams11}) for future work.

%
\vspace*{-0.7cm}
\section{Structures beyond the local volume}\label{resfar}

\begin{figure*}
  \centering  \includegraphics[height=0.23\textwidth]{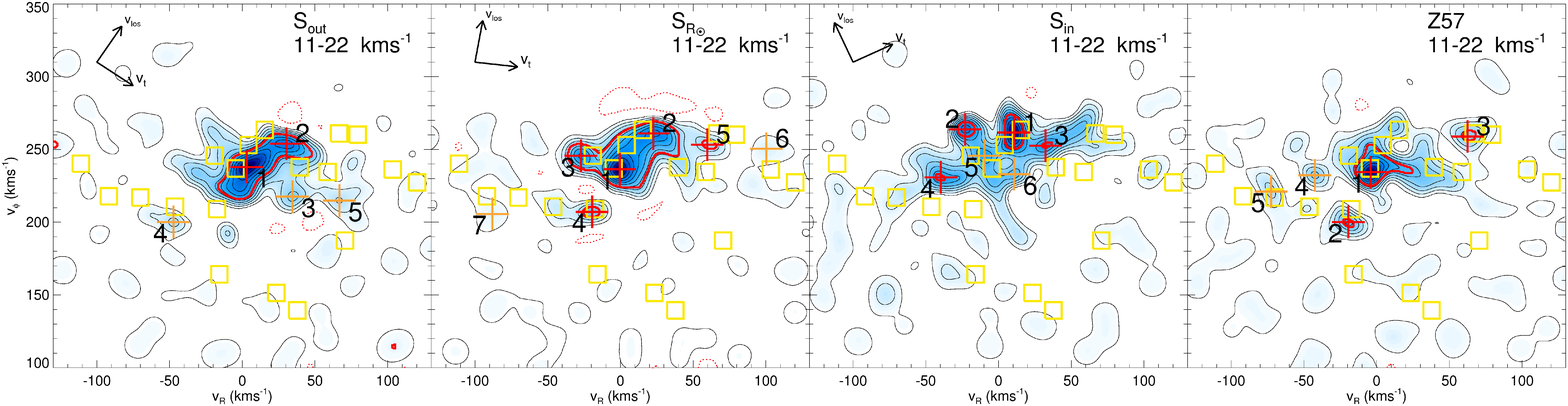}
  \caption{Velocity structures at scales of $11$-$22\kms$ in cylindrical velocities for the subsamples S$_{\rm{out}}$ (outside Solar circle), {S$_{\rm{R\sun}}$} (Solar circle), S$_{\rm{in}}$ (inside Solar circle), and Z57 ($-700\leq Z\leq-500\pc$). Yellow squares show the positions of the local sample (LS) groups. Colour code is the same as in Fig. \ref{f:UV}b.}
  \label{f:UVss}%
\end{figure*}

The three first panels of Fig.~\ref{f:UVss} show the velocity structures of the subsamples S$_{\rm{out}}$, {S$_{\rm{R\sun}}$} and S$_{\rm{in}}$. Now we use a Galactic cylindrical coordinate system, $v_R$ and $v_\phi$, which is more suitable for distant regions. We use Local Standard of Rest (LSR) circular velocity of $v_0=247\kms$ and Solar movement with respect to the LSR of $(U_\sun,V_\sun)=(10.0,11.0)\kms$ (2nd model in Table 2 of \citealt{McMillan10}). For convenience, we take $v_R$ and $v_\phi$ to be positive towards the Galactic Centre and towards rotation (clockwise), respectively, to get the same orientation as $U$--$V$. Larger WT scales ($11$-$22\kms$) are considered now due to the larger velocity errors in these subsamples (Table \ref{t:subsamples}). We find that for {S$_{\rm{R\sun}}$} and S$_{\rm{out}}$ the general shape of the distribution is preserved and some groups show a relative configuration similar to the LS. Although we show here only three distant regions, the evolution in the location and intensity of the main peaks as one moves away from the LS to these volumes is found to be smooth and continuous. For example, the most populated group in {S$_{\rm{R\sun}}$} (group 1 in the second panel) is the Pleiades group. Also groups 2 and 3 are in the Sirius and Hyades kinematic positions, respectively. Group 4 is the Hercules stream and group 5 corresponds to $\gamma$Leo. We also see the Wolf elongation, but with no associated maxima. Similar identifications can be done for S$_{\rm{out}}$. By contrast, the velocity distribution in S$_{\rm{in}}$ looks quite different to the LS, its overall shape being elongated with a positive slope in the $v_R$ direction. There are several central groups but they are distributed along the general elongated structure and do not show the same relative configuration as for LS, {S$_{\rm{R\sun}}$} and S$_{\rm{out}}$.

We also find that some groups change their location on the velocity plane in distant regions. The yellow squares in Fig.~\ref{f:UVss} indicate the cylindrical velocities of the LS groups. For instance, Hercules (group 4) in S$_{\rm{R\sun}}$ is at the same cylindrical velocity as locally (peak inside the square), but in S$_{\rm{out}}$ it has moved to lower $v_\phi$. We see that this transition is continuously traced when we consider samples in small steps advancing from S$_{\rm{out}}$ to {S$_{\rm{R\sun}}$}, and progressively to inner Galacocentric radius. For S$_{\rm{in}}$, Hercules does not appear as a group separated from the central structures. The $v_\phi$ of the Wolf elongation follows a similar behaviour with radius. We also observe that some groups move in $v_R$. For instance, Sirius (group 2) shifts to the right of its corresponding square (the one at highest $V$) for S$_{\rm{out}}$. 

The arrows in the upper left of each plot show the median directions of the radial ($v_{los}$) and transverse ($v_t$) velocity. For these regions, as the radial velocity errors are small ($e_{v_{los}}\sim2\kms$), the errors affect mainly the transverse motion ($e_{v_t}\sim17\kms$), distorting the groups in that direction. For the case of S$_{\rm{out}}$, the velocity errors or distance biases would never move Hercules in the observed way. By contrast, for S$_{\rm{in}}$ a distance underestimation could shift Hercules closer to the centre of the distribution. However, as showed in \citet{Burnett11}, distances for giants may have a slight bias but in the opposite direction.

We have evaluated the effects of neglecting the extinction in deriving distances for stars in these regions. By using the dust maps of \citet{Schlegel98}, we find an upper limit on the extinction in the $J$-band of 0.2 $mag$ (by considering extinction at infinity) for the three subsamples, which would introduce a distance error $<10\%$, which is small compared to the $28\%$ uncertainty due to the distance derivation method itself. We can also test that our results do not depend on the distance method by using only red clump stars. We select these by $J-K$ colour and gravity using the criteria in \citet{Siebert11a} and re-derive the distance from their $J$ or $K$ magnitude assuming absolute magnitudes $M_J=-0.87$ and $M_K=-1.55$ \citep{Zwitter08}. This gives us subsamples with $30$-$40\%$ of the initial number of stars. Considering a dispersion of $20\%$ in the absolute magnitude for this population, we find that the transverse velocity errors are reduced substantially, from $e_{v_t}\sim17\kms$ to $e_{v_t}\sim11\kms$. Nonetheless, with smaller velocity errors and alternative distances, our results do not change.
 
We perform another test to establish if errors only could explain the differences in the distant velocity distributions. We study if these velocity distributions could belong to the same parent distribution as the LS, given the velocity errors. To do this, we incorporate the observational errors of {S$_{\rm{R\sun}}$}, S$_{\rm{in}}$ and S$_{\rm{out}}$ in the LS velocity distribution. In practice, we use the proper motions and radial velocities of the LS but the distances and sky positions of the distant subsamples to generate 5000 simulated velocity distributions after error convolution. Then we compare the mock distributions with the original one (at S$_{\rm{in}}$, S$_{\rm{out}}$, or {S$_{\rm{R\sun}}$}), by computing the distribution of the Kolmogorov-Smirnov (KS) 2d statistic. Instead of using this statistic directly, as a reference we also compute the distribution of the KS statistic between the mock subsamples in a given location themselves. The area of overlap of the two KS distributions (obtained for the mock-original and for the mock-mock comparisons) thus gives us the probability that the distant velocity distributions belong to the same parent distribution as the LS, given the velocity errors. Notice that our mock samples may have too large errors because first, the distance errors may be overestimated \citep{Burnett11}, and second, the LS velocities have already their own error. We find that for the three subsamples, this probability is $<0.06\%$. Repeating the same test but introducing also a distance bias of $\pm10,\,20,\,50$ $\%$ does not increase the probability. Only for S$_{\rm{in}}$ we find a $5\%$ probability of consistency with the LS, given the velocity errors, if the distances were overestimated by $10$ or $20\%$. But in this case, this bias would move Hercules far from the centre of the velocity distribution, contrary to what we see.

Fig. \ref{f:UVss} (right panel) shows the velocity distribution of the subsample Z57 (median height below the plane of $Z=-0.6\kpc$). Again, some of the LS groups can still be recognized and are \3s significant. For instance, Hercules is here group 2, the Wolf elongation is present and group 3 
 has the same velocity as $\gamma$Leo. For a sample in the same cylinder as Z57 but with heights in the range $-800\leq Z\leq-600\pc$ (median of $Z=-0.7\kpc$) Hercules and the equivalent to $\gamma$Leo are still detected but with \2s confidence.
\vspace*{-0.5cm}

\section{Summary and discussion}

We have analysed, for the first time, the velocity distributions in regions that are $\sim1\kpc$ far from the Sun using RAVE data and a structure detection technique based on the wavelet transform. We have found that most of the main local kinematic groups (Hyades-Pleiades, Sirius, Hercules, Wolf elongation, $\gamma$Leo) can be still detected at $\sim1\kpc$ from the Sun in the direction of anti-rotation. Some of them also persist at $700\pc$ below the Galactic plane. Therefore, these groups are actually large scale features. Given also the spread in age and metallicity of some of these groups (A08), this supports a dynamical origin instead of being dispersed clusters. 

In the distant disc volume at the Solar circle the groups appear approximately at the same cylindrical velocities as locally. In contrast, for the distant regions (in the direction of anti-rotation) at different Galactocentric radii ($\pm600\pc$) the groups appear shifted in the velocity plane. These are the first significant changes in the kinematic structures ever observed in distant disc velocity distributions. Especially at inner radii the relative configuration between the groups changes considerably. These trends are consistent with the typical spatial scale of variation of the kinematic structures in models of e.g. spiral arm effects (A11).

In particular, we show that the Hercules group moves to larger (respectively, smaller) azimuthal velocity at inner (respectively, outer) Galactocentric radius, while it appears at the same velocity at the Solar radius. This is the behaviour expected for this structure if caused by the bar's Outer Lindblad resonance (fig.~4 in \citealt{Dehnen00}), although it can be also explained based on the epicycle theory for a resonant feature \citep{Quillen11}. We also observe shifts in Galactocentric radial velocity, which is also observed in several models (A11, \citealt{Quillen11, Minchev11}).

We have also found a new significant group in the Solar neighbourhood and confirmed the presence of structures that were not significant in other studies and samples. Some of the observed features such as the Wolf elongation are generally only vaguely considered in the modelling. These and especially the observed groups in distant regions may constitute the keys for the understanding of the complex kinematics of the MW disc. In particular, the magnitude and direction of the velocity shifts detected here in distant regions may tell us about the specific mechanism that produces them (bar or spiral arm related). Quantitative comparison with the models could allow us to constrain the properties of the spiral arms and the bar of the MW, breaking the degeneracy found when fitting the models only to the main local groups.

\vspace*{-0.5cm}

\section*{Acknowledgments}
{\small
TA and AH acknowledge funding support from the European Research
Council under ERC-StG grant GALACTICA-24027. Funding for RAVE has been provided by: the Australian Astronomical Observatory; the Leibniz-Institut f\"ur Astrophysik Potsdam (AIP); the Australian National University; the Australian Research Council; the French National Research Agency; the German Research Foundation (SPP 1177 and SFB 881); the European Research Council (ERC-StG 240271 Galactica); the Istituto Nazionale di Astrofisica at Padova; The Johns Hopkins University; the National Science Foundation of the USA (AST-0908326); the W. M. Keck foundation; the Macquarie University; the Netherlands Research School for Astronomy; the Natural Sciences and Engineering Research Council of Canada; the Slovenian Research Agency; the Swiss National Science Foundation; the Science \& Technology Facilities Council of the UK; Opticon; Strasbourg Observatory; and the Universities of Groningen, Heidelberg and Sydney. The RAVE web site is at http://www.rave-survey.org.}

\vspace*{-0.1cm}

\bibliographystyle{mn2e2} 

\bibliography{mybib} 
\label{lastpage}

\end{document}